\documentclass[aps,prb,twocolumn,showpacs]{revtex4}
\usepackage{graphicx}

\begin{document}
\title{Dynamical transitions and sliding friction of the phase-field-crystal
model with pinning }

\author{J.A.P. Ramos $^{1,2}$, E. Granato $^{2,3}$, S.C. Ying $^3$, C.V. Achim $^{4,5}$,
K.R. Elder $^6$, and T. Ala-Nissila $^{3,4}$}

\address{$^1$ Departamento de Ci\^encias Exatas, Universidade
Estadual do Sudoeste da Bahia, 45000-000 Vit\'oria da Conquista,
BA,Brasil}
\address{$^2$Laborat\'orio Associado de Sensores e Materiais,
Instituto Nacional de Pesquisas Espaciais,12245-970 S\~ao Jos\'e dos
Campos, SP, Brazil}
\address{$^3$Department of Physics, P.O. Box 1843, Brown University,
Providence, RI 02912-1843, USA}
\address{$^4$Department of Applied Physics and COMP Center of Excellence,
P.O. Box 1100, Helsinki University of Technology, FI-02015 TKK,
Espoo, Finland}
\address{$^5$
Institut f\"{u}r Theoretische Physik II: Weiche Materie,
Heinrich-Heine-Universit\"{a}t D\"{u}sseldorf,
Universit\"{a}tsstra{\ss}e 1, D-40225 D\"{u}sseldorf, Germany}
\address{$^6$Department of Physics, Oakland University, Rochester,
Michigan 48309-4487, USA}


\begin{abstract}
We study the nonlinear driven response and sliding friction behavior
of the phase-field-crystal (PFC) model with pinning including both
thermal fluctuations and inertial effects. The model provides a
continuous description of adsorbed layers on a substrate under the
action of an external driving force at finite temperatures, allowing
for both elastic and plastic deformations. We derive general
stochastic dynamical equations for the particle and momentum
densities including both  thermal fluctuations and inertial effects.
The resulting coupled equations for the PFC model are studied
numerically. At sufficiently low temperatures we find that the
velocity response of an initially pinned commensurate layer shows
hysteresis with dynamical melting and freezing transitions for
increasing and decreasing applied forces at different critical
values. The main features of the nonlinear response in  the PFC
model are similar to the results obtained previously with molecular
dynamics simulations of particle models for adsorbed layers.

\end{abstract}


\maketitle

\section{Introduction}

In recent years, considerable attention has been given to the study
of driven adsorbed layers in relation to sliding friction phenomena
between surfaces at the microscopic level
\cite{Israelechvili,Book,Conf,Persson93,Granato99,Granato00}. The
nonlinear response of the adsorbed layer is a central problem for
understanding experiments on sliding friction between two surfaces
with a lubricant \cite{Israelechvili} or between adsorbed layers and
an oscillating substrate \cite{Krim,Mistura}. Various elastic and
particle models have been used to study the driven dynamical
transitions and the sliding friction of adsorbed monolayers
\cite{Persson93,Book,Conf,Granato99}. A fundamental issue in
modeling such systems is the origin of the hysteresis and the
dynamical melting and freezing transitions associated with the
different static and kinetic frictional forces when increasing the
driving force from zero and decreasing from a large value,
respectively. In driven lattice systems, hysteresis can occur in the
underdamped regime where inertial effects are present or in the
presence of topological defects such as dislocations and thermal
fluctuations \cite{Fisher85}. Topological defects can be
automatically included in a full microscopic model involving
interacting atoms in the presence of a substrate potential using
realistic interaction potentials.   However, numerical observation of
the full complexities of the phenomena is severely limited by the
small system sizes that can be studied, even when
simple Lennard-Jones potentials are employed \cite{Persson93}.

Recently, a phase-field-crystal (PFC) model was introduced
\cite{Elder02,Elder04,Elder07} that allows for both elastic and
plastic deformations within a continuous description of the particle
density while still retaining information on atomic length scales.
By extending the PFC model to take into account the effect of an
external pinning potential \cite{achim06}, a two-dimensional version
of the model has been used to describe commensurate-incommensurate
transitions in the presence of thermal fluctuations \cite{ramos08}
and the driven response of pinned lattice systems without thermal
fluctuations and inertial effects.
\cite{achim09}. 
However, in order to study fully the sliding friction behavior,
\emph{both} inertial effects \emph{and} thermal fluctuations need to
be taken into account.

In this work, we study sliding friction of an adsorbed layer
via the nonlinear response of the PFC model to an external force in
the presence of a pinning potential.
To this end, we first derive the general stochastic dynamical
equations for the particle and momentum density fields taking full
account of the thermal fluctuations and inertial effects. The
resulting coupled equations are studied numerically. At low
temperatures, we find that the velocity response of an initially
commensurate layer shows hysteresis with dynamical melting and
freezing transitions for increasing and decreasing applied forces at
different critical values. The main features of the nonlinear
response are similar to the results obtained previously with
molecular dynamics simulations of particle models. However, the
dynamical melting and freezing mechanisms are significantly
different. In the PFC model, nucleation occurs via stripes rather
than closed domains found in the particles model.

\section{Dynamics of the PFC model with inertial effects}

In the generalized PFC model, the system is represented by a coarse
grained effective Hamiltonian that is a functional of the number
density field. To take into account inertial effects in the
dynamics, we need to consider  the contribution of the kinetic
energy to the total energy of the system in addition to the
configurational energy.
Thus we consider the momentum density, $\vec g(\vec x)=\rho(\vec x)
\vec v(x)$, a dynamical variable in the coarse grained Hamiltonian
in addition to the particle density field $\rho(\vec x)$. The total
effective Hamiltonian in the presence of an external force $\vec f$
can be written as
\begin{equation}
H_t = H_{kin}+ H_{int} - \int d \vec x \rho(\vec x) \vec x \cdot  \vec f ,
\label{energy}
\end{equation}
where $H_{kin}$ is the kinetic energy contribution given by
\begin{equation}
H_{kin}= \int d \vec x \frac{{\vec g}^2 (\vec x)}{2 \rho(\vec x)},
\end{equation}
and $H_{int}(\rho)$ is the configurational contribution to the
effective Hamiltonian in the original PFC model. The last term is
due to the presence of the the external force $\vec f$.

In the absence of energy dissipation, the time dependence of $\rho$
and $\vec g$ are determined by the Poisson brackets  $\{ H_t,\rho
\}$ and $\{ H_t,\vec g \}$. At finite temperatures, additional
dissipative noise terms are present in the dynamical equations. The
noise satisfies the fluctuation dissipation relations
\cite{chaikin,mazenkob} which allows the system to reach thermal
equilibrium in the absence of external perturbations.
We include the  dissipative noise term directly in the dynamical
equations for $\vec g$, which is a nonconserved field.

For the particle density $\rho(\vec x,t)$ we have

\begin{equation}
\frac{\partial \rho}{\partial t}  = -\sum_{j}\int d {\vec x}^\prime
\{\rho(\vec x),g_j(\vec x^\prime)\} \frac{\delta H_t}{\delta
g_j(\vec x^\prime)}, \label{rho}
\end{equation}
and for the momentum density $\vec g (x,t)$
\begin{eqnarray}
\frac{\partial g_i}{\partial t}  &=& - \int d {\vec x}^\prime
\{g_i(\vec x),\rho(\vec x^\prime)\} \frac{\delta H_t}{\delta
\rho(\vec x^\prime)}  \cr & &-\sum_{j} \int d {\vec x}^\prime
\{g_i(\vec x),g_j(\vec x^\prime)\} \frac{\delta H_t}{\delta g_j(\vec
x^\prime)} \cr & &- \eta \frac{\delta H_t}{\delta g_i} + \nu_i(\vec
x,t), \label{g1}
\end{eqnarray}
where $\eta$ is a dissipative coefficient and the noise $\vec \nu
(\vec x,t)$ has variance
\begin{equation}
\langle \nu_i(\vec x,t) \nu_j(\vec x^\prime, t^\prime) \rangle =
2 k_B T \eta \delta (\vec x - \vec x^\prime) \delta (t -
t^\prime)\delta_{i,j}.
\end{equation}
The Poisson brackets for the mass and momentum densities are given
by \cite{chaikin,mazenkob}
\begin{eqnarray} \label{poisson}
\{\rho(\vec x),g_i(\vec x^\prime)\}&=& \nabla_i(\rho(\vec
x)\delta(\vec x-\vec x^\prime)); \\ \nonumber
 \{g_i(\vec x),\rho(\vec x^\prime)\}&=&
\rho(\vec x)\nabla_i\delta(\vec x-\vec x^\prime); \\
\nonumber \{g_i(\vec x),g_j(\vec x^\prime)\}&=&\nabla_j(g_i(\vec
x)\delta(\vec x-\vec x^\prime)) - \nabla_i^{\prime}(g_j(\vec
x)\delta(\vec x-\vec x^\prime)).
\end{eqnarray}
Substituting Eqs. (\ref{poisson}) into Eqs. (\ref{rho}) and
(\ref{g1}) gives
\begin{equation}
\frac{\partial \rho}{\partial t}  =  -\nabla \cdot \vec g;
\label{rho2}
\end{equation}
\begin{eqnarray} \label{g2}
 \frac{\partial g_i}{\partial
t}= -\rho\nabla_i  \frac{\delta H_{int}}{\delta \rho} &-& \eta
\frac{g_i}{\rho} + \rho f_i + \nu_i(\vec x,t) \\
\nonumber & & -\sum_j \nabla_j \frac{g_i g_j}{\rho},
\end{eqnarray}
for a spatially uniform external force ($f_i$).
We can redefine the coefficient $\eta \rightarrow \rho \eta$ to
remove the denominator from the term $g_i/\rho$ in Eq. (\ref{g2}).
With this change the variance of the noise $\vec \nu(x,t)$ becomes
\begin{equation}
\label{g2b} \langle \nu_i(\vec x,t) \nu_j(\vec x^\prime, t^\prime)
\rangle = 2 k_B T \eta \rho(\vec x) \delta (\vec x - \vec
x^\prime) \delta (t - t^\prime)\delta_{i,j}
\end{equation}

To leading order in the momentum density $\vec g$, we drop the
quadratic term in Eq. (\ref{g2}) giving
\begin{equation}
\frac{\partial \rho}{\partial t}  =  -\nabla \cdot \vec g;
\label{rho3}
\end{equation}
\begin{equation}
 \frac{\partial g_i}{\partial
t}= -\rho\nabla_i  \frac{\delta H_{int}}{\delta \rho} + \rho f_i -
\eta g_i + \nu_i(\vec x,t). \label{g3}
\end{equation}
Similar dynamical equations for PFC models with internal
dissipation were obtained in Ref. \onlinecite{sami}.
If the effective Hamiltonian $H_{int}$ is known, then these coupled
stochastic dynamical equations should provide a full description of
the particle and momentum densities in presence of fluctuations
represented by the noise $\nu_i(x,t)$ with correlations proportional
to the temperature $T$ and inertial effects determined by the
damping parameter $\eta$. In the overdamped limit,
$\partial{g}/\partial{t} =0$, with $T=0$ and $f=0$, the equation for
the time evolution of $\rho$ obtained by inserting $\vec g$ from Eq.
(\ref{g3}) in Eq. (\ref{rho3}) reduces to the deterministic equation
for the density which has been obtained from classical density
functional theory of liquids \cite{Voigt}.

\section{PFC model with pinning and thermal fluctuations}

In the presence of  an external pinning potential
\cite{achim06,ramos08,achim09}, a specific form of the
configurational energy contribution $H_{int}$ to the
total effective Hamiltonian in Eq.(\ref{energy})
has been
proposed
which is an extension
of the standard PFC model free energy functional used in many
applications \cite{Elder04,Elder07}. In dimensionless form, this
effective interaction Hamiltonian  $H_{int}=H_{\rm pfc}$ can be written
as
\begin{equation}
 H_{\rm pfc}  =  \int d \vec x \{  \frac{1}{2} \psi[r+(1+\nabla^2)^2] \psi +
\frac{\psi^4}{4} + V \psi \}, \label{cpfcp}
\end{equation}
where $\psi(\vec x)$ is a continuous field, $V(\vec x)$ represents
the external pinning potential and $r$ is a parameter. The phase
field $\psi(\vec x)$ can be regarded as a measure of deviations of
the particle number density $\rho(\vec x)$ from a uniform reference value
$\rho_0$, such that $\psi(\vec x)= (\rho(\vec x) -\rho_0)/\rho_0$.
It is a conserved field and its average value, $\bar{\psi}$,
represents another parameter in the model. The
intrinsic wave vector of the model $\vec k_i$  has no preferred
directions and its magnitude is set to unity in the present work.

In the absence of a pinning potential, the Hamiltonian of Eq.
(\ref{cpfcp}) is minimized by a configuration of the field
$\psi(\vec x)$ forming a hexagonal pattern of peaks with reciprocal
lattice vectors of magnitude $|\vec k_h| \approx 1$, in an
appropriate range of values for the parameters $r$ and $ \bar{\psi}$
in the model. This periodic structures of peaks in $\psi$ can be
regarded as a simple model of an atomic layer.

%

To study the nonlinear dynamical behavior we take the form of
$H_{int}$ given by Eq.(\ref{cpfcp}) together with the kinetic energy
and the external force terms for the total effective Hamiltonian
$H_t$ in the dynamical equations Eq. (\ref{g3}) and in Eq.
(\ref{g2b}). We make the additional simplifying approximation that $
\rho(\vec x) \approx \rho_0$ in Eq. (\ref{g2b}) and in the
coefficient of the first term in Eq. (\ref{g3}). This approximation
ensures that the effective diffusion coefficient in the model is
positive definite for any temperature and driving force. Another
motivation for this approximation is to show that the dynamical
equations used in the previous works \cite{achim06,achim09} follows
from the more general Eqs. (\ref{rho3}) and (\ref{g3}). Setting
$\rho_0=1$, we obtain,
\begin{eqnarray}\label{cpeqs}
\frac{\partial \psi}{\partial t}  =  -\nabla \cdot \vec g; & &\\
\nonumber \frac{\partial g_i}{\partial t}= -\nabla_i
\frac{\delta H_{pfc}}{\delta \psi} &+& \psi f_i - \eta g_i + \nu_i(\vec x,t); \\
\nonumber \langle \nu_i(\vec x,t) \nu_j(\vec x^\prime, t^\prime)
\rangle &=& 2 k_B T \eta \delta (\vec x - \vec x^\prime) \delta (t
- t^\prime)\delta_{i,j}.
\end{eqnarray}
Here we have redefined $\vec g \rightarrow \vec g  + \rho_0 \vec
f/\eta$ to remove a uniform term on right hand side of the
equation for $\vec g$.

The above coupled equations can be combined in a single equation by
applying the operator $\nabla \cdot$ to both sides of the second
equation and using the first one to eliminate $\nabla \cdot \vec
g$, giving
\begin{eqnarray}\label{sgeqs}
\frac{\partial^2 \psi}{\partial t^2} + \eta \frac{\partial
\psi}{\partial t}  &=& \nabla^2
\frac{\delta H_{\rm pfc}}{\delta \psi}  - \vec f \cdot \nabla \psi + \xi(\vec x,t); \\
\nonumber
 \langle \xi(\vec x,t) \xi(\vec x^\prime, t^\prime) \rangle &=&
2 k_B T \eta \nabla^2 \delta (\vec x - \vec x^\prime) \delta (t -
t^\prime).
\end{eqnarray}
When the driving force $\vec f$ and the external pinning potential
are  set to zero, the dynamical equation  above is identical to the
one used in Ref. \onlinecite{stefano} to study propagating density
modes in the PFC model. It can also be obtained by introducing
inertial effects in the dynamical equation of the PFC model through
a memory function of exponential form\cite{Galenko}. In the limit of
large $\eta$ when $\partial g_i/\partial t$ in Eqs. (\ref{cpeqs})
or, equivalently, $\partial^2 \psi/\partial t^2$ in Eqs.
(\ref{sgeqs}), can be neglected, these equations reduce to the
familiar overdamped dynamical equations used in the previous works
\cite{achim06,achim09} without inertial effects at zero temperature.

\section{Numerical results and Discussion}

In this section, we present our numerical results for  the velocity
response of the PFC model in presence of an external pinning
potential under a uniform applied force. For the numerical
calculations, the phase field $\psi(\vec x)$ and momentum density
field $\vec g(\vec x)$ are defined on a space square grid with $d x
= dy = \pi /4$ with periodic boundary conditions. System sizes $L
\times L$ with $L=64$ to $128$ where used. The Laplacians and
gradients were evaluated using finite differences.

\begin{figure}
\includegraphics[ bb= 1cm 1cm  19cm   27cm, width=7.5 cm]{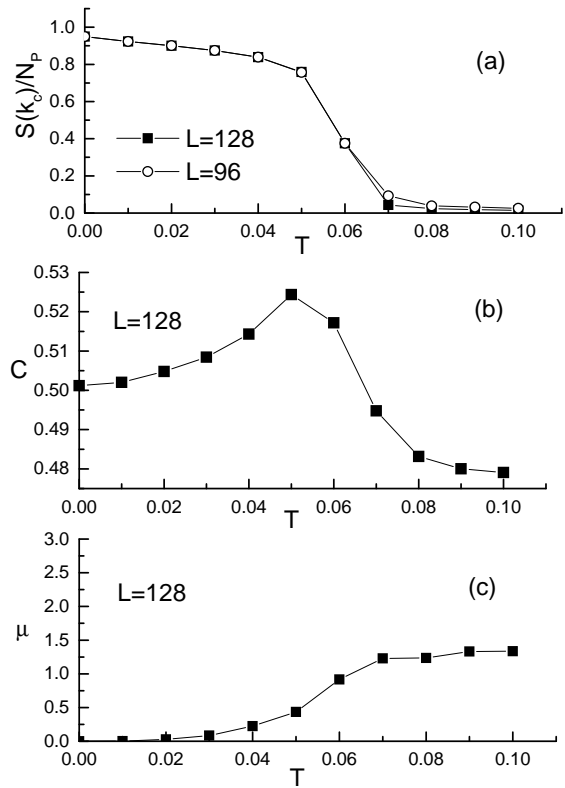}
\caption{ Temperature dependence of the (a) scaled structure-factor
peak $S(k_c)/N_p$; (b) specific heat $C$, and mobility $\mu$ for the
model without an external driving force.}\label{mob}
\end{figure}

We consider a pinning potential $V(\vec x)$ representing a substrate
with square symmetry
\begin{equation}
V(\vec x)=-V_0[\cos(k_0 x)+ \cos(k_0 y)],
\end{equation}
where $k_0 $ defines the period of the pinning potential for both
the $x$ and $y$ directions. The lattice misfit between the
phase-field crystal and the pinning potential can be defined as
$\delta_{\rm m}=(1-k_0)$. We choose the parameters of the PFC model
as $r =-0.25$, $\bar \psi = -0.25$, a lattice mismatch $\delta=-0.5$
and a pinning strength $V_0=0.15$. For these parameters, the
ground-state configuration in the absence of an external driving
force is a $c(2 \times 2)$ phase \cite{ramos08,achim09}, where every
second site of the lattice of the pinning potential corresponds to a
peak in the phase field $\psi(\vec{r})$. This commensurate
configuration is stable below a commensurate melting temperature
$T_c \approx 0.055$. This is determined from the temperature
dependence  of the structure factor as well as that of the specific
heat. The structure factor $S(\vec k)$ is calculated from the the
positions $\vec R_j$ of the peaks in the phase field as
\begin{equation}
S(\vec k) = \langle \sum_{j,j'=1}^{N_P}\frac{1}{N_P}e^{-i\vec k
\cdot (\vec R_j -\vec R_{j'})} \rangle . \label{sfe}
\end{equation}
Here, $\langle ... \rangle$ denotes a time average which is
equivalent to a thermal average at equilibrium. For increasing
temperatures thermal fluctuations disorder the layer and the scaled
structure factor $S(\vec k_c)/N_p$ evaluated at the primary
reciprocal lattice vector  $\vec k_c $ of the commensurate phase
decreases from a value of unity at $T=0$ rapidly through $T_c$ to
zero at higher temperatures. The transition is broadened due to
finite size effects as shown in Fig. \ref{mob}(a). Another signature
of this commensurate melting transition is that the specific heat
develops a broad peak near $T_c$ corresponding to an increase
in the fluctuations in $\psi$ at the transition as shown in Fig.
\ref{mob}(b). Finally, the mobility $\mu$ defined as $ \lim_{f \to
0} (v/f)$ where $v$ is the drift velocity, also changes
qualitatively through the transition. In the commensurate phase, a
finite threshold for the sliding of the overlayer exists and hence
the mobility is vanishingly small.  As the system melts above
$T_c$
the mobility rapidly rises and
reaches a plateau at higher temperatures.
This behavior is shown in
Fig. \ref{mob}c.
For the study of the nonlinear response and sliding
friction of the system, we focus on an initial state which
corresponds to a well pinned  $c(2 \times 2)$ phase initially,
corresponding to $T=0.01 \ll T_c$, and a damping parameter $\eta
=0.4$.

In our numerical calculations, a  driving force along the $x$
direction is increased from zero to a maximum value above a critical
depinning force  and then decreased back to zero. For each value of
the force, the coupled equations (\ref{cpeqs}) are solved using an
Euler algorithm with time step $dt=0.001 - 0.005$. The implicit
trapezoidal method was also used to check numerical stability. For
each value of the force,  $10^6$ time steps were used to allow the
system to reach a velocity steady state and an additional period
with the same number of time steps were used to calculate the
average velocity and other time averaged quantities.

To study the velocity response of the PFC model, we need to
determine the velocity of the peaks \cite{ramos08} in the phase
field $\psi(\vec x)$. This is done by determining the time
dependence of the peak positions ($\vec R_i(t)$) in $\psi$.
The steady state drift velocity $\vec v$ for the system
is obtained from the peak velocities $\vec v_i$ ($d\vec R_i/dt$) as
\begin{equation}
\vec v = \langle \frac{1}{N_P}\sum_{i=1}^{N_P} \vec v_i(t) \rangle,
\label{velpeak}
\end{equation}
where $N_P$ is the number of peaks and $\langle ... \rangle$ denotes
time average.
The steady state structure factor $S(\vec k)$  which is a measure of
translational order, is also calculated from the the peak positions
$\vec R_j$, as in Eq.(\ref{sfe}).

\begin{figure}
\includegraphics[ bb= 3cm 3cm 19cm 15cm, width=7.5 cm]{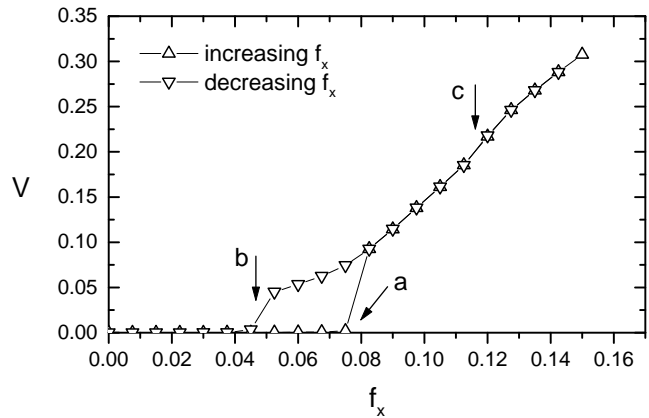}
\caption{Velocity response as a function of applied force. Arrows
correspond to critical values $f_a$, $f_b$ and $f_c$.  }\label{vxf}
\end{figure}

The most notable qualitative feature of the velocity response to the
driving force $\vec f$ (as shown in Fig. \ref{vxf}) is that it shows
hysteresis behavior with two different critical force thresholds $
f_a \approx 0.075$ for increasing forces and  $f_b \approx 0.045$
for decreasing forces. These two  threshold values correspond to
the static frictional force and kinetic frictional force
respectively. As the force is increased beyond $f_a$, the velocity
jumps abruptly from zero to a finite value whereas when the force is
decreased below $f_b$ the velocity of the sliding layer drops
abruptly and become pinned by the external potential to form an
immobile commensurate state again. Below, we present the microscopic
details of the configurations for the system in the neighborhood of
the two thresholds. This allows us to characterize the change in
velocity response at $f_a$ as a dynamical force induced melting
transition of the initial commensurate state,  and the second
transition at $f_b$ as a dynamical force induced freezing of the
sliding phase.

To study these transitions we first examine the behavior of the
steady state structure factor, as shown in Fig. \ref{sf}. We focus
on the dependence of  $S(Q)$ on the driving force, where $\vec Q$ is
the dominant reciprocal lattice vector of the layer. $\vec Q = \vec
k_c$ corresponds to the primary reciprocal lattice vector for the
$c(2\times 2)$ phase and $\vec Q= \vec k_h$ to the reciprocal
lattice vector of the hexagonal phase in absence of the driving
force. Consistent with the velocity response behavior in Fig.
\ref{vxf}, on increasing the force beyond $f_a$, $S(k_c)$ drops
abruptly to zero. This is the onset of the dynamical force induced
melting transition of the initial $c(2x2)$ commensurate state. The
behavior of $S( k_c)$ is analogous to the temperature induced
disordering transition shown in Fig. \ref{mob}(a). On decreasing the
force, the value of $S( k_c)$ stays vanishingly small until the
force drops below the threshold $f_b$, at which point $S(k_c)$
rapidly increases to a value corresponding to the commensurate
pinned state.
The other interesting feature is that for $ f> f_c \approx 0.12$,
the structure factor shows clear $6$-fold coordinated peaks at the
reciprocal lattice vectors $\vec Q = \vec k_h$ corresponding to a
hexagonal phase, which grows as the driving force increases. This
implies that at a driving force larger than this third threshold
$f_c$, there is another dynamic continuous transition from a
disordered phase into an incommensurate hexagonal phase. This state
emerges as the average effect of the external pinning potential
becomes less and less important at high sliding velocities and the
steady state then approximately corresponds to the phase-field
crystal in the absence of the pinning potential which has hexagonal
symmetry in the equilibrium state.
However, since the scaled structure factor for the hexagonal phase
is still much less than unity, this incommensurate hexagonal phase
is not fully ordered even at at the largest force values ($f<0.15$)
studied so far.

\begin{figure}
\includegraphics[ bb= 3cm 4cm 19cm 16cm, width=7.5 cm]{Structure2.eps}
\caption{Scaled structure-factor peak $S(Q)/N_p$ as a function of
applied force. Here $\vec Q $ stands for the primary reciprocal
lattice vector for either the $c(2 \times 2)$ phase ($\vec k_c$) or
the hexagonal phase ($\vec k_h $). Filled and open symbols
correspond to increasing and decreasing forces,
respectively.}\label{sf}
\end{figure}

\begin{figure}
\includegraphics[ bb= 3cm 1cm 19cm 14cm, width=7.5 cm]{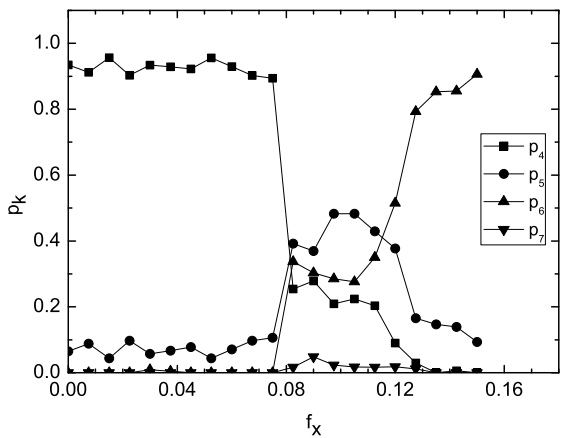}
\caption{ Fraction of density peaks  $p_k$ with coordination number
$k$ (4,5,6 and 7 nearest neighbors) for increasing applied force.
}\label{coord}
\end{figure}

\begin{figure}
\includegraphics[ bb= 3cm 1cm 19cm 14cm, width=7.5 cm]{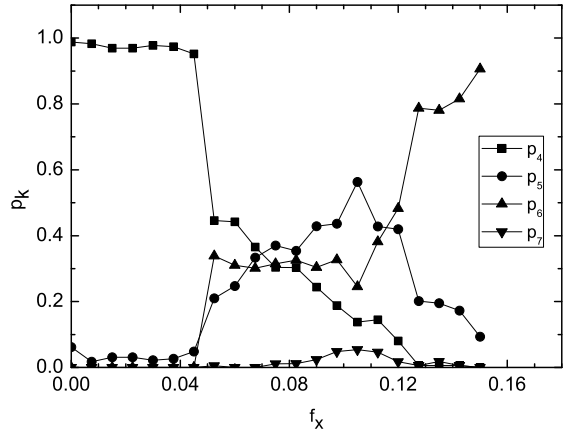}
\caption{ Fraction of density peaks  $p_k$ with coordination number
$k$ for decreasing applied force.}\label{coord2}
\end{figure}

\begin{figure}
\includegraphics[ bb= 3cm 9cm 16.5cm 19cm, width=7.5 cm]{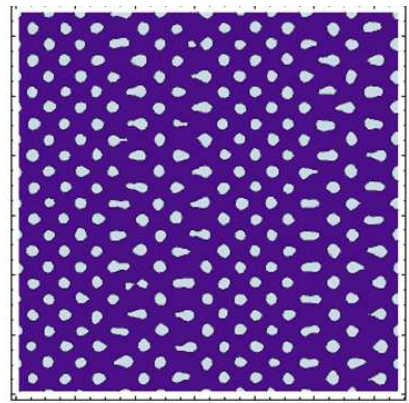}
\caption{Snapshot of the density field in the sliding state at
$f_x=0.0525$.}\label{config}
\end{figure}

\begin{figure}
\includegraphics[ bb= 2.5cm 9cm 17.5cm 20cm, width=7.5 cm]{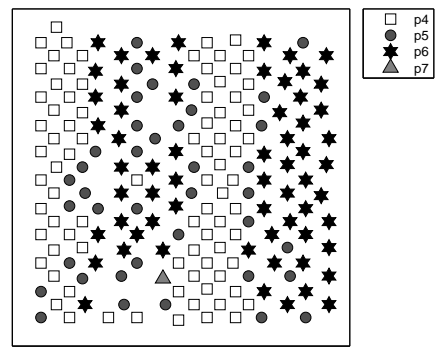}
\caption{ Configuration of the density peak locations with
corresponding coordination numbers for the density plot in Fig.
\ref{config}.}\label{stripes}
\end{figure}

To better understand the dynamic melting transition at $f_a$ and the
freezing transition at $f_b$, we inspect the steady state
configurations near the two thresholds as well as the region between
the two thresholds in real space, which yield  more direct and
detailed  information than the structure factor at the peak values.
Treating the peaks in the phase field as "particles", the
coordination number of each particle in  the commensurate
$c(2\times2)$ phase  is $4$ while for the incommensurate hexagonal
phase it is $6$.  Furthermore,  the core of dislocations in a
hexagonal lattice correspond to particles with $5$-fold and $7$-fold
coordination numbers. Thus, further insight into the microscopic
nature of the steady state configurations can be obtained by
inspecting the fraction of particles with coordination numbers
$p_4$, $p_5$,  $p_6$, $p_7$. The results are shown in Figs.
\ref{coord} and Fig. \ref{coord2} for increasing and decreasing
applied forces respectively. Fig. \ref{coord} shows that when the
force $f$ is increased beyond $f_a$, besides the rapid drop in $p_4$
consistent with the structure factor data, the fraction of other
coordination numbers  $p_5$,  $p_6$, $p_7$ also increases
significantly untill $f$  reaches $f_c$, beyond which $p_5$ and
$p_7$ start to decrease and we have a continuous transition to an
incommensurate hexagonal phase. Thus the nature of steady state
above $f_a$ is a strongly disordered state analogous to the high
temperature phase in the absence of driving force above the
commensurate melting temperature. On decreasing the force below
$f_a$, the data in Fig. \ref{coord2} shows that the system remains
in a melted state with large disorder until the threshold $f_b$ is
reached, below which we have only $4$-fold coordination number and
the system returns to a pinned $c(2\times2)$ phase. Taken together,
the qualitative behavior of the structure factor and the
coordination number strongly suggest that the transitions at $f_a$
and $f_b$ can be regarded as a force induced dynamical melting and
freezing transition respectively.

\begin{figure}
\includegraphics[ bb= 3.5cm 14cm 16.5cm 28cm, width=7.5 cm]{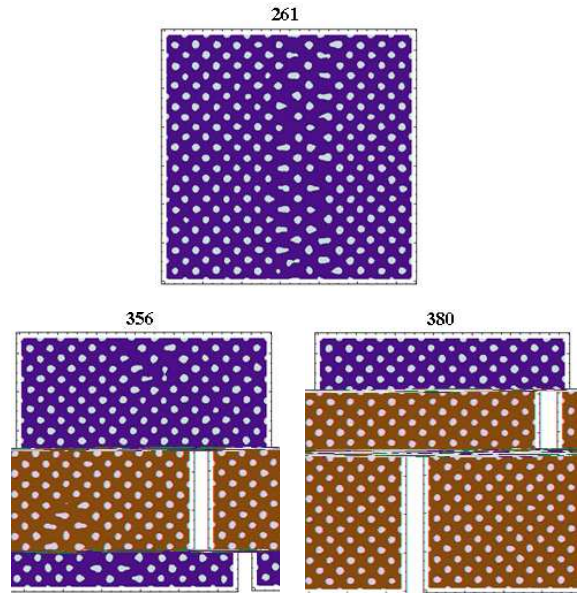}
\caption{Snapshots of the density field for increasing times after
starting from the moving state of Fig. \ref{config} and reaching the
pinned commensurate state, at a force  $f_x=0.045$ just below $f_b$.
Times are in units of $10^3$ time steps.} \label{rep}
\end{figure}

Finally, we look at a snapshot of the phase field $\psi(x)$ obtained
in the steady state for a driving force just above the dynamical
freezing threshold $f_b$. This is shown in Fig. \ref{config}. It
consists of stripes of commensurate $c(2\times2)$ phase separated by
disordered domain walls. These domain walls are mobile liquid like
regions as confirmed by a count of the fraction of various
coordination numbers shown in  Fig. \ref{stripes}. The commensurate
stripes have only $4$-fold coordination numbers, while in the liquid
like domain walls, there is a mixture of $5$-fold, $6$-fold and
$7$-fold coordination numbers. Starting from this steady state which
still has a non-zero average sliding velocity, we can watch the
dynamical freezing in real time as the driving force is reduced to a
value just below the freezing threshold $f_b$. The time sequence of
the freezing is shown in  Fig. \ref{rep}. In returning to the pinned
state, the domain wall regions gradually shrink and eventually
disappear.


The main features of the dynamical melting and freezing  and
hysteresis effects in PFC model described above are similar to those
found earlier by molecular dynamics simulations of particle models
with interacting Lennard-Jones potentials \cite{Persson93} and the
uniaxial Frenkel-Kontorova model \cite{Granato99} under a driving
force starting from a commensurate state. The similarity of results
from these very different models demonstrates the universality of
the hysteresis loop for models with inertia effects and the
macroscopic consequence of stick and slip motion. When compared
with the Lennard-Jones particle model one notable difference is the
mechanism of the dynamical freezing at $f_b$. In the present PFC
model, it involves parallel liquid-like domain walls similar to the
uniaxial Frenkel-Kontorova model, although in the absence of the
driving force there is no easy direction. In the Lennard-Jones
model, nucleation and growth occur via closed pinned domains
\cite{Persson93}. The origin of this intriguing difference requires
further investigation of both atomistic and PFC models. One
possibility is that the nucleation of stripes is related to the
absence of a fixed constraint on the number of peaks in the PFC
model. In this case, the mechanism of the dynamical freezing found
in the present PFC model should be compared to the results of
particle models with a constant chemical potential rather than with
a fixed particle number. Unfortunately, such results are currently
unavailable.

\section{Conclusions}
In this paper, we have  derived general stochastic dynamical
equations for the particle and momentum density fields including
both  thermal fluctuations and inertial effects. The new equations
are applied to the study of the nonlinear response to an external
driving force for a PFC model with a pinning potential. The model
describes a driven  adsorbed layer as a continuous density field,
allowing for elastic and plastic deformations. The numerical results
showed that at low temperatures, the velocity response of an
initially commensurate layer shows hysteresis with dynamical melting
and freezing transitions for increasing and decreasing applied
forces at different critical values. The inclusion of both thermal
fluctuations and inertial effects  are crucial for a correct
description of these dynamical transitions. The main features of the
nonlinear response, in particular the hysteresis loop separating the
static friction and sliding friction thresholds are similar to the
results obtained previously with particle models. However, the
details of the dynamical melting and freezing mechanisms are
significantly different. In the PFC model considered here, they
correspond to nucleation of stripes rather than closed domains found
in particle models. It is possible to describe more realistic
sliding adsorbed systems if the parameters of the model are adjusted
to match experimental systems, similar to the recent works for the
colloidal systems \cite{Voigt} and Fe \cite{akusti}.

\begin{acknowledgments}

J.A.P.R. acknowledges the support from Secretaria da Administra\c
c\~ao do Estado da Bahia. E.G. was supported by Funda\c c\~ao de
Amparo \`a Pesquisa do Estado de S\~ao Paulo - FAPESP (Grant No.
07/08492-9). S.C.Y. also acknowledges FAPESP (Grant No. 09/01942-4)
for supporting a visit to Instituto Nacional de Pesquisas Espaciais.
This work has been supported in part also by the Academy of Finland
through its COMP Center of Excellence grant and by joint funding
under EU STREP 016447 MagDot and NSF DMR Award No. 0502737.   K. R.
E. acknowledges support from NSF under Grant No.  DMR-0906676. E.G.
thanks Sami Majaniemi for many helpful discussions.

\end{acknowledgments}

\end{document}